\documentclass[aps, prd, amsmath, amssymb, preprintnumbers, groupedaddress, superscriptaddress, twocolumn, nofootinbib, 10pt]{revtex4-1}

\usepackage[hidelinks]{hyperref}
\usepackage{xcolor}

\makeatletter
\renewcommand{\p@subsection}{}
\renewcommand{\p@subsubsection}{}

\newcommand{\KB}{\overline{\cal K}}
\newcommand{\bbZ}{{\mathbb{Z}}}
\newcommand{\bbP}{{\mathbb{P}}}

\newcommand{\cS}{{\cal S}}

\makeatother
\hypersetup{
    pdftitle={A Quadrillion Standard Models from F-theory},
    pdfauthor={Mirjam Cvetic, James Halverson, Ling Lin, Muyang Liu, Jiahua Tian},
    pdfsubject={String Landscape, Standard Model, Chiral spectrum, F-theory}
}

\begin{document}

\preprint{\texttt{UPR-1297-T}}

\title{A Quadrillion Standard Models from F-theory}
\author{Mirjam Cveti\v{c}} \affiliation{Department of Physics and Astronomy, University of Pennsylvania, Philadelphia, PA 19104-6396, USA}
  \affiliation{Center for Applied Mathematics and Theoretical Physics, University of Maribor, Maribor, Slovenia}
\author{James Halverson} \affiliation{Department of Physics, Northeastern University, Boston, MA 02115-5000, USA}
\author{Ling Lin} \affiliation{Department of Physics and Astronomy, University of Pennsylvania, Philadelphia, PA 19104-6396, USA}
\author{Muyang Liu} \affiliation{Department of Physics and Astronomy, University of Pennsylvania, Philadelphia, PA 19104-6396, USA}
\author{Jiahua Tian} \affiliation{Department of Physics, Northeastern University, Boston, MA 02115-5000, USA}

\begin{abstract}
\noindent We present ${\cal O}(10^{15})$ string compactifications with the exact chiral spectrum of the Standard Model of particle physics. 
This ensemble of globally consistent F-theory compactifications automatically realizes gauge coupling unification.
Utilizing the power of algebraic geometry, all global consistency conditions can be reduced to a single criterion on the base of the underlying elliptically fibered Calabi--Yau fourfolds.
For toric bases, this criterion only depends on an associated
polytope and is satisfied for at least ${\cal O}(10^{15})$
bases, each of which defines a distinct compactification.
\end{abstract}

\maketitle

\parskip 2pt plus 1pt minus 1pt

\interfootnotelinepenalty=10000


\section{Introduction and Summary}
\label{sec:intro}

As a theory of quantum gravity that naturally gives
rise to rich gauge sectors at low energies, 
string theory is a leading candidate for a unified theory.
Achieving unification is an ambitious goal that requires
accounting for all aspects of our physical world,
which includes not only a rich cosmological history, but
also the detailed structure of the Standard Model of particle physics.

In this paper we present an explicit construction that
guarantees the existence of ${\cal O}(10^{15})$ fully consistent
string compactifications which realize the exact chiral particle spectrum
of the minimally supersymmetric Standard Model (MSSM).
This construction is performed
in the framework of F-theory \cite{Vafa:1996xn}, a strongly coupled generalization
of type IIB superstring theory.
It captures the non-perturbative back-reactions of 7-branes onto the compactification space $B_3$ in terms of an elliptically fibered Calabi--Yau fourfold $\pi: Y_4 \rightarrow B_3$ over it.
Gauge symmetries, charged matter, and Yukawa couplings are then encoded beautifully by $Y_4$'s singularity structures in codimensions one, two, and three, respectively.\footnote{We refer the interested reader to \cite{Weigand:2018rez,*Cvetic:2018bni} and references therein for recent reviews on F-theory.}

In the present work, we consider a class of elliptically fibered Calabi--Yau fourfolds giving rise to precisely the three-generation MSSM spectrum provided certain geometric conditions on the base of the fibration are satisfied.
We perform a concrete analysis, finding ${\cal O}(10^{15})$ such bases.
All these models come equipped with moduli-dependent quark and lepton Yukawa couplings, as well as gauge coupling unification at the compactification scale.

The existence of a very large number of Standard Model realizations in string theory
could perhaps be anticipated within the set of an even larger number of string compactifications
(see, e.g., \cite{Ashok:2003gk,*Taylor:2015xtz,*Halverson:2017ffz}) that form the so-called string landscape.
Indeed, though Standard Model realizations within the landscape could potentially be scarce \cite{Blumenhagen:2004xx, *Gmeiner:2005vz}, recent works hint towards an astronomical number of them \cite{Anderson:2012yf, *Anderson:2013xka,*Constantin:2018xkj}.
Our construction \textit{explicitly} demonstrates this possibility, increasing the number of concretely known, global Standard Model compactifications in string theory by about ten orders of magnitude.

There are also explicit constructions of the Standard Model in other corners of string theory.
Some of the early examples of globally consistent intersecting brane models \cite{Cvetic:2001nr,*Cvetic:2001tj} in type II compactifications (see also \cite{Blumenhagen:2005mu} and references therein) were strongly constrained by global consistency conditions such as tadpole cancellation.
In the heterotic string, the typical difficulties of constructions like \cite{Braun:2005ux,*Braun:2005nv,Bouchard:2005ag,*Bouchard:2006dn,Anderson:2007nc,*Anderson:2009mh} arise from having a stable hidden bundle and the existence of Yukawa couplings.
These issues are solved elegantly in F-theory through the geometrization of non-perturbative stringy effects:
(almost all\footnote{In F-theory, D3-tadpole cancellation requires extra care, and will be a major theme in our constructions.}) global conditions analogous to tadpole cancellation or bundle stability are automatically taken care of by having a compact, elliptic Calabi--Yau fourfold $Y_4$, and the presence
or absence of Yukawa couplings can be easily read off from codimension three singularities of $Y_4$.

Despite these advantages, only a handful \cite{Cvetic:2015txa,Cvetic:2018ryq} of F-theory compactifications that realize the exact chiral spectrum of the MSSM are currently known, due to focusing on a very simple base, $B_3 = \bbP^3$.
This limitation will be avoided in the current work
by instead studying smooth toric varieties,
which provide a much larger class \cite{Halverson:2016tve,*Carifio:2017bov,Altman:2018zlc} of geometries.
To take advantage of this large ensemble, we first
construct a class of elliptic fibrations (based on the class
$\bbP_{F_{11}}$ in \cite{Klevers:2014bqa})
that can be consistently fibered over all 
such toric threefolds.

Every such fibration realizes the precise Standard Model gauge group $[SU(3) \times SU(2) \times U(1)]/\bbZ_6$ as well as its matter representations and Yukawa couplings \cite{Klevers:2014bqa,Cvetic:2015txa,Cvetic:2017epq}.
Moreover, all models exhibit gauge coupling unification at the compactification scale, compatible with the existence of a complex structure deformation to a geometry realizing the Pati--Salam model with unified gauge coupling \cite{Klevers:2014bqa,Cvetic:2015txa}.

Furthermore, for each compatible $B_3$ there exists a $G_4$-flux that induces three families of chiral fermions.
These models have a particularly pleasant feature: all global consistency conditions on the flux (including quantization and D3-tadpole cancellation) can be reduced to a single criterion on the intersection number $\KB^3$ of the anti-canonical class $\KB$ of the base $B_3$.
For toric threefolds which have a description in terms of a reflexive polytope $\Delta$, $\KB^3$ depends only on the point configuration of $\Delta$ and not its
triangulation.
On the other hand, for a single polytope there can be multiple different toric threefolds associated with the different fine regular star triangulations (FRSTs) of $\Delta$, the number of which grows exponentially with the number of lattice points in the polytope \cite{Halverson:2016tve}.
Putting together these different components,
we find that the number $N_\text{SM}^\text{toric}$ of globally consistent three-family Standard Models in our construction 
is
\begin{align}
	7.6 \times 10^{13}\lesssim N^\text{toric}_{\text{SM}}\lesssim 1.6 \times 10^{16} .
\end{align}
We emphasize that this number is construction dependent; F-theory could realize more Standard Models.

The detailed derivation of this count first requires the construction in section \ref{sec:universal_fibration} of a class of elliptic fibrations with a flux inducing three chiral families.
All flux consistency conditions reduce to a single criterion on the base $B_3$.
To count how many $B_3$ satisfy this criterion, we discuss the methods to construct FRSTs of 3D polytopes in section \ref{sec:counting}, which ultimately lead us to ${\cal O}(10^{15})$ possibilities.
We close in section \ref{sec:discussion} with some
geometric and physical comments, as well as future directions.

\section{Universally Consistent Fibrations with Three Families}
\label{sec:universal_fibration}


We now present our construction of three-family Standard Model vacua; see
appendix \ref{app:resolution} for further details.

For that, we first consider an elliptic curve that is a specialized cubic inside $\mathbb{P}^2$ with homogeneous coordinates $[u:v:w]$, given by the vanishing of the polynomial
\begin{align}\label{eq:hypersurface_polynomial}
\begin{split}
	P :=  s_1 u^3 + s_2 u^2 v + s_3 u v^2 + s_5 u^2 w + s_6 u v w + s_9 v w^2.
\end{split}
\end{align}
By promoting the coefficients $s_i$ to rational functions over a K\"ahler threefold $B_3$, one obtains a singular, elliptically fibered fourfold $\pi: Y^{(s)}_4 \rightarrow B_3$.
For $Y_4^{(s)}$ to be Calabi--Yau, the functions $s_i$ have to be holomorphic sections of line bundles on $B_3$ with first Chern classes $[s_i] \in H^{1,1}(B_3,\bbZ)$ given by \cite{Klevers:2014bqa,Cvetic:2015txa}:
\begin{align}\label{eq:coefficients_F11_classes}
	\begin{array}{lll}
		[s_1] = 3\,\KB - \cS_7 - \cS_9 \, , &  [s_2] = 2\,\KB - \cS_9 \, , & [s_6] = \KB \, , \\ [1ex]
		[s_3] = \KB + \cS_7 - \cS_9 \, , & [s_5]= 2\,\KB - \cS_7 \, , &  [s_9] = \cS_9 \, ,
	\end{array}
\end{align}
where $\KB \equiv c_1(B_3)$ is the anti-canonical class of $B_3$.
The classes $\cS_{7,9} \in H^{1,1}(B_3, \bbZ)$ parametrize different fibrations over the same base, on which
$\{s_i=0\}$ define effective divisors.

When all $s_i$ are generic, (that is, irreducible and $s_i \neq s_j$ for $i\neq j$), F-theory compactified on $Y_4^{(s)}$ has the gauge symmetry $[SU(3) \times SU(2) \times U(1)]/\mathbb{Z}_6$ \cite{Klevers:2014bqa, Cvetic:2017epq}.
The global gauge group structure is reflected in the precise agreement between the geometrically realized matter representations and those of the Standard Model:
\begin{align}\label{eq:matter_reps}
		({\bf 3,2})_{\frac16} \, , \, \, \, ({\bf 1,2})_{-\frac12} \, , \, \, \, ({\overline{\bf 3}, {\bf 1}})_{-\frac23} \, , \, \, \, ({\overline{\bf 3}, {\bf 1}})_{\frac13} \, , \, \, \, ({\bf 1,1})_{1} \, .
\end{align}
These data can be extracted via the M-/F-theory duality from an explicit resolution $Y_4$ of $Y_4^{(s)}$, which preserves the Calabi--Yau structure (see appendix \ref{app:resolution}).


A chiral spectrum in F-theory requires a non-zero flux $G_4 \in H^{2,2}(Y_4)$, which must also be specified.
For the relevant subspace of so-called primary vertical $G_4$-fluxes, there is by now a large arsenal of computational methods \cite{Lin:2015qsa,Lin:2016vus} (see also \cite{Cvetic:2013uta,Cvetic:2015txa,MayorgaPena:2017eda}) that allows us to determine \textit{base-independently} the most general flux on $Y_4$.


For physical consistency, this $G_4$-flux has to satisfy certain 
conditions.
The first condition is a proper quantization \cite{Dasgupta:1999ss,*Witten:1996md,Collinucci:2010gz}:
\begin{align}\label{eq:quantization_condition}
	G_4 + \frac12 \, c_2(Y_4) \in H^{2,2}(Y_4, \mathbb{Z}) \, ,
\end{align}
where $c_2(Y_4)$ is the second Chern class of $Y_4$.
Heuristically, this condition ensures that the notion of fermions (that requires a flux-dependent spin structure on subspaces of $Y_4$) is well-defined.
Since explicitly verifying this condition for concrete geometries is difficult,
we will content ourselves with the usual necessary consistency checks \cite{Intriligator:2012ue,Cvetic:2013uta,Cvetic:2015txa,Lin:2016vus,Cvetic:2018ryq}.
The second consistency condition is a D3-tadpole
satisfying
\cite{Sethi:1996es},
\begin{align}\label{eq:D3-tadpole_general}
	n_\text{D3} = \frac{\chi(Y_4)}{24} - \frac{1}{2} \int_{Y_4} G_4 \wedge G_4  \stackrel{!}{\in} \mathbb{Z}_{\geq 0} \, .
\end{align}
While integrality follows as a consequence of the quantization condition \eqref{eq:quantization_condition}, positivity aids in ensuring the stability of the compactification.

We must also impose phenomenological constraints on the
flux.
A massless electroweak
hypercharge $U(1)_Y$ 
is guaranteed if the D-terms vanish \cite{Grimm:2010ks,Grimm:2011tb}:
\begin{align}\label{eq:U1-D-term}
\begin{split}
	\forall \eta \in H^{1,1}(B_3) : \quad \int_{Y_4} G_4 \wedge \sigma \wedge \pi^*\eta \stackrel{!}{=} 0 .
\end{split}
\end{align}
Here, $\sigma$ is the $(1,1)$-form Poincar\'{e}-dual to the so-called \textit{Shioda-divisor} associated with the $U(1)$ \cite{Park:2011ji, *Morrison:2012ei}.
A three-family chiral Standard Model requires that \cite{Donagi:2008ca,*Braun:2011zm,*Marsano:2011hv,*Krause:2011xj,*Grimm:2011fx},
\begin{align}\label{eq:chirality_formula}
	\chi({\bf R}) = \int_{\gamma_{\bf R}} G_4 \stackrel{!}{=} 3 \, ,
\end{align}
for all representations ${\bf R}$ in \eqref{eq:matter_reps}.
The geometric data $c_2(Y_4)$, $\chi(Y_4)$, and the matter surfaces $\gamma_{\bf R}$ were computed in \cite{Klevers:2014bqa,Cvetic:2015txa}.
In the Appendix, we provide the explicit expression of the generic vertical flux in the resolution $Y_4$ presented in \cite{Klevers:2014bqa} and explain in detail how the above conditions can be checked using well-studied topological methods.


We now present our main result, on how these consistency conditions can be satisfied for a large ensemble of explicit geometries.
For that, we first consider the flux configuration \eqref{eq:generic_flux} on (smooth) fibrations $Y_4$ with $\cS_{7,9} = \KB$, which simplifies the expressions for the topological quantities \eqref{eq:D3-tadpole_general}--\eqref{eq:chirality_formula}. 
In fact, one can show that all consistency conditions are reduced to a \textit{single} criterion on $B_3$ from the D3-tadpole:
\begin{align}\label{eq:tadpole_simplified}
	n_\text{D3} = 12 + \frac{5}{8}\,\KB^3 - \frac{45}{2\,\KB^3} \stackrel{!}{\in} \bbZ_{\geq 0} \, ,
\end{align}
where $\KB^3$ denotes the triple self-intersection number of the anti-canonical class $\KB$ of the base.
This dramatic simplification only requires $\KB^3$ of appropriate value and a base that allows irreducible and distinct $s_i$, all of which are sections of the anti-canonical class.

In summary, we have constructed a class of elliptically fibered Calabi--Yau fourfolds which gives rise in F-theory to the Standard Model gauge group and matter representations with three chiral generations.
The only consistency requirement that guarantees flux quantization and D3-tadpole cancellation is that the base $B_3$ of the fibration is a smooth K\"ahler threefold with non-rigid irreducible anti-canonical divisors that satisfy \eqref{eq:tadpole_simplified}.
In fact, 
the only values $\KB^3$ can take such that $n_\text{D3} \in \mathbb{Z}_{\geq 0}$ are
\begin{align}\label{eq:final_condition_on_KB}
	\KB^3 \in \{2, 6, 10, 18, 30, 90\} \, .
\end{align}




\section{Counting Standard Model Geometries}
\label{sec:counting}

Any smooth threefold $B_3$ with non-rigid anti-canonical divisors satisfying \eqref{eq:final_condition_on_KB} realizes a globally consistent three-family MSSM in F-theory.
A subset of such spaces, which can be enumerated combinatorially, is the set of weak Fano toric threefolds encoded by 3D reflexive polytopes $\Delta$.
While there are ``only'' 4319 such polytopes \cite{Kreuzer:1998vb}, each $\Delta$ can specify inequivalent manifolds $B_3$ through different \textit{fine-regular-star triangulations} (FRSTs) of the polytope, whose numbers can be very large \cite{Halverson:2016tve}.

What makes this ensemble particularly attractive for our purpose is the fact that the intersection number $\KB^3$ is determined solely by the polytope $\Delta$, and is completely triangulation-independent.
Therefore any $B_3$ associated to an FRST of
$\Delta$ gives rise to a consistent chiral three-generation 
MSSM by our construction, provided that the triangulation-independent constraint on $\KB^3$
is satisfied.
In fact, there is a set $S$ with 708 polytopes that satisfy \eqref{eq:final_condition_on_KB}. By our
construction we immediately have
\begin{align}
N^\text{toric}_{\text{SM}} = \sum_{\Delta \in S} N_{\text{FRST}}(\Delta) \, ,
\end{align}
where $N_{\text{FRST}}(\Delta)$ is the number of FRSTs of $\Delta$.

Hence, the problem of counting the number of consistent F-theory models that admit the chiral MSSM spectrum by
our construction reduces to counting FRSTs of reflexive polytopes.

Since $N_{\text{FRST}}(\Delta)$ grows exponentially with the number of lattice points in $\Delta$, the set of consistent threefolds $B_3$ is dominated by triangulations of the largest polytope \cite{Halverson:2016tve}, labelled $\Delta_8$ in the list of \cite{Kreuzer:1998vb}.
The FRSTs of this polytope (with $\KB^3 = 6$ and 39 lattice points) cannot be all constructed explicitly using the standard computer programs such as \texttt{SageMath} \cite{sage}.
To enumerate them, we therefore follow the strategy put forward in \cite{Halverson:2016tve} to provide bounds on $N_{\text{FRST}}(\Delta_8)$.

The idea is to reduce the complexity by first counting the number of \textit{fine-regular triangulations} (FRTs) of each \textit{facet} of a polytope $\Delta$.
Since the facets are two dimensional polytopes, it is possible to use brute-force on the combinatorics of FRTs for (almost\footnote{For facets with more than 15 lattice points, brute-forcing FRTs also becomes computationally too costly.
For these facets, we use different methods outlined in \cite{Halverson:2016tve} to obtain lower and upper bounds for the number of FRTs.
}) all polytopes' facets.
By virtue of the reflexivity of $\Delta$, any combination of FRTs of all its facets yields fine star triangulation
of $\Delta$.

The drawback of this approach is that the triangulation of $\Delta_8$ obtained this way is not guaranteed to be regular.
To tackle this issue, we randomly pick $1.3 \times 10^4$ samples out of ${\cal O}(10^9)$ fine-star triangulations constructed by gluing together FRTs of the facets $\Delta_8$.
Out of these samples, we find roughly $\frac23$ to be also regular triangulations.
Combining the factor $\frac23$ with the bounds of \textit{fine-star triangulations} (FSTs) for $\Delta_8$ \cite{Halverson:2016tve}, we then obtain $2.6 \times 10^{13} \leq N_{\text{FRST}}(\Delta_8) \leq 1.6 \times 10^{16}$.

For the other polytopes in $S$ (i.e., those leading to threefolds satisfying \eqref{eq:final_condition_on_KB}) we can either compute all FRSTs, or we can resort to a similar estimation as with $\Delta_8$ if the
polytope is too large to use brute force all FRSTs.
We find that these other polytopes sum up to ``only'' $\sim 5 \times 10^{13}$ FRSTs, which confirms the dominance of $\Delta_8$.
In total, we therefore expect the number of consistent three-family F-theory Standard Models in our
construction over toric threefold bases to be
\begin{align*}
	7.6 \times 10^{13}\lesssim N^\text{toric}_{\text{SM}}\lesssim 1.6 \times 10^{16} \, .
\end{align*}

\section{Discussion and Outlook}\label{sec:discussion}

We have presented a construction that ensures the
existence of ${\cal O}(10^{15})$ explicit, globally
consistent string compactifications having the
exact chiral spectrum of the Standard Model within the framework of F-theory. To our
knowledge, this is the largest such ensemble in the
literature, outnumbering existing results by about
$10$ orders of magnitude.
The models arise by varying the base of one ``universal'' class of elliptic fibrations introduced in \cite{Klevers:2014bqa,Cvetic:2015txa}.
We have only focused on the set of toric bases, which already produces around a quadrillion examples.
However, we expect that the ensemble of Standard Models arising from our construction is of orders of magnitude larger than this, as might be shown, for instance, by including non-toric bases.

All these models have in common that the Higgs and lepton doublets are localized on the same matter curve.
As such, this curve must have non-zero genus to allow for the existence of vector-like pairs \cite{Bies:2014sra,*Bies:2017fam}.
Given the homology class of the doublet curve \cite{Cvetic:2015txa} and our restriction $\cS_{7,9} = \KB$, the genus in question is indeed $g= 1 + 9/2\KB^3 > 0$, since $\KB^3\geq 2$ by \eqref{eq:final_condition_on_KB}.
It would be very interesting, albeit extremely difficult with current methods, to study the precise complex structure dependence of the number of Higgs doublets and other charged vector-like pairs in this ensemble.

Furthermore, since our models have no additional (possibly massive) abelian gauge symmetries, all Yukawa couplings relevant for the Standard Model are automatically realized perturbatively, as
can be shown by an explicit study of codimension
three singularities \cite{Klevers:2014bqa}.
However, this in turn also implies that certain proton decay operators compatible with the Standard Model gauge group will in general be present \cite{Cvetic:2015txa}.
We expect that in some corners of the moduli space, which incidentally could also support high-scale supersymmetry breaking, these operators can be suppressed.
Another avenue could be to instead focus on ``F-theory Standard Models'' that have additional ($U(1)$ \cite{Lin:2014qga,Lin:2016vus} or R-parity \cite{Cvetic:2018ryq}) selection rules, and estimate their numbers in the toric base landscape.
We leave this for future work.

One interesting aspect of our ensemble is gauge coupling unification without a manifest GUT-origin at the compactification scale.
It can be easily read off geometrically from the divisors on $B_3$, which the 7-branes supporting the gauge symmetries in the type IIB picture wrap.
Due to our restriction $\cS_{7,9}=\KB$, both $SU(3)$ and $SU(2)$ gauge symmetries are realized on anti-canonical divisors $\{s_9=0\}$ and $\{s_3=0\}$ with class $\KB$.
Therefore, the gauge couplings are $g_{3,2}^2 = 2/\text{vol}(\KB)$ \cite{Grimm:2010ks,Bonetti:2011mw}.\footnote{
The factor of 2 arises because in F-theory, the normalization dictated by the geometry is one where the Cartan generators satisfy $\text{tr}_\text{fund}(T_i\,T_j) = C_{ij}$ with $C$ the Cartan matrix.
On the other hand, the particle physics convention necessary to determine the coupling is $\text{tr}_\text{fund}(T_i\,T_j) = \frac{\delta_{ij}}{2}$.
}
The $U(1)_Y$ coupling is the inverse volume of the so-called height-pairing divisor $b \subset B_3$ \cite{Cvetic:2012xn}, which has been computed in \cite{Klevers:2014bqa,Cvetic:2001tj} and reduces to $b = 5\KB/6$ in our ensemble.
Therefore, we have the standard MSSM gauge coupling unification,
\begin{align}\label{eq:gauge_unification}
	g_3^2 = g_2^2 = \frac53\,g_Y^2 = \frac{2}{\text{vol}(\KB)} \, ,
\end{align}
which for our models is achieved at the compactification scale.
While this scale as well as the actual values of the couplings will depend on the details of moduli stabilization, the relationship \eqref{eq:gauge_unification} is independent of K\"ahler moduli.
It would be interesting to see if this relationship originates from an honest geometric realization of a GUT-structure.
Given the known connection of our ensemble to a Pati--Salam $[SU(4)\times SU(2)^2]/\bbZ_2$ model \cite{Klevers:2014bqa,Cvetic:2015txa}, we expect an underlying $SO(10)$.

Our results may provide
phenomenological motivation for the study of new moduli
stabilization scenarios.
Specifically, though gauge
coupling unification is automatic in our ensemble, it is
natural to ask whether the correct value
$\alpha_{\text{GUT}}\simeq.03$ can be obtained
in canonical moduli stabilization schemes.
For instance, the KKLT and Large Volume scenarios \cite{Kachru:2003aw,*Balasubramanian:2005zx} assume that cycles are at sufficiently large
volume to safely ignore string worldsheet instanton
corrections to the K\" ahler potential. 
This is essential because
it is not known how to systematically compute and control all instanton contributions
in $\mathcal{N}=1$ backgrounds. 
A necessary condition
for safely ignoring these corrections is to have
$\text{vol}(C)>1$ (in string units) for all
curves $C \subset B_3$. This condition defines a stretched out subset of the
K\" ahler cone \cite{Demirtas:2018akl}, where it was also shown that the
K\" ahler cones become increasingly narrow for increasing
$h^{1,1}$.
In effect, this forces toric divisors to be increasingly
large in order to safely ignore
worldsheet instantons, leading to smaller gauge couplings, because on toric $B_3$ the class $\KB$ is the sum of all toric divisors.
Brief calculations suggest that the correct $\alpha_{\text{GUT}}$ cannot be obtained in this controlled regime, in which case realistic 
models in our scenario are not consistent with the KKLT
or LVS scenarios. Firmly concluding this requires
a more detailed study, but we emphasize that 
it would not
rule out our models, and instead motivate the study
of new moduli stabilization scenarios that allow
for the observed value of gauge couplings. 

Our compactifications
also exhibit D3-branes. 
These sectors generically give
rise to $U(1)$ gauge theories that could be cosmologically
relevant as dark photons. 
Each has
its own open string moduli, the position of the D3-brane,
which are massless at tree level but may be stabilized
by non-perturbative effects due to their appearance
in instanton prefactors \cite{Ganor:1996pe,*Baumann:2006th}. However, since all but
one of the toric divisors are rigid in the geometries we study, 
it is likely that
there are many instanton corrections to the 
superpotential. 
Each instanton acts with an attractive
force on the D3-brane, pulling it toward the associated
divisor, but the existence of many such contributions
would provide competing forces that stabilize the D3-brane
away from each toric divisor. 
In particular, due to
these competing effects we see a priori no reason that the
D3-branes should be stabilized anywhere near the
$SU(3)$ or $SU(2)$ 7-branes, in which case jointly
charged matter in the form of $3$-$7$ strings decouple
from the spectrum. Such a scenario gives rise to 
numerous dark photon sectors that have cosmological
effects only through kinetic mixing with the visible
sector and with one another. It would be interesting
to study these sectors further, in light of 
current and future dark photon experiments.

We note that gravity cannot be decoupled in our ensemble
since the Standard Model gauge divisors are in the anti-canonical class,
yielding a non-trivial
interplay between gravity and the visible sector.
This interplay arises due to the details of our
construction and could lead to other interesting interactions between particle physics and cosmology.
At the level of
toric geometry, the models of our ensemble differ from one another
by how the facets are triangulated.
This does not affect the structure of the anti-canonical divisors that realize $SU(3)$ and
$SU(2)$, and thus the particle physics is relatively
insensitive to details of the triangulation; it is,
after all, what gives rise to the large number of
Standard Models in our construction. The triangulation
is critical, however, for moduli stabilization. For instance, the classical
K\"ahler potential on K\"ahler moduli
is determined by triangulation-dependent
topological intersections. This affects numerous
aspects of the cosmology of these models,
including inflation.

This visible sector universality in the midst 
of cosmological diversity might lead one to question
the extent to which these should be counted as truly
different models. Though a natural question, it has
a clear answer: since the geometries are different they
lead to distinct four-dimensional effective theories
below the Kaluza--Klein scale, each of which could give
rise to numerous metastable vacua. Instead, our view
is that the universal structure in the visible sector
provides some evidence for a long-held hope in the string
landscape, that, despite large numbers of vacua, there
could exist semi-universal features that lead
to meaningful statistical predictions.

\begin{acknowledgments}
\noindent We thank Craig Lawrie and Cody Long for helpful discussions and Jonathan Carifio for providing the dataset of \cite{Altman:2018zlc}.
MC and LL are supported by DOE Award DE-SC0013528.
MC further acknowledges the support by the Fay R.~and Eugene L.~Langberg Endowed Chair and the Slovenian Research Agency.
JH is supported by NSF Grant PHY-1620526.
\end{acknowledgments}

\appendix

\section{Technical Aspects of the Construction}\label{app:resolution}

While the physics is fully determined by the singular elliptic fibration $\pi^{(s)}: Y_4^{(s)} \rightarrow B_3$, extracting the relevant information is most easily done in a crepant resolution, i.e., a resolution of the singularities which preserve the Calabi--Yau condition \cite{Vafa:1996xn,Weigand:2018rez,Cvetic:2018bni}.
One crepant resolution $\pi: Y_4 \rightarrow B_3$ of the singular fibration $Y^{(s)}$ obtained from the specialized cubic \eqref{eq:hypersurface_polynomial} can be conveniently described by toric blow-ups on the ambient $\mathbb{P}^2$.
This introduces exceptional divisors $\{e_i = 0\}$, $i=1,...,4$, whose coordinates $e_i$ are subject to additional scaling relations amongst each other and the $\mathbb{P}^2$ coordinates $[u:v:w]$.
These relations, together with intersection properties, are encoded in a reflexive 2d polygon (labelled $F_{11}$ in \cite{Klevers:2014bqa}), which also determines the polynomial,
\begin{align}
\begin{split}
	P := &  s_1\,e_1^2\,e_2^2\,e_3\,e_4^4\,u^3 + s_2\,e_1\,e_2^2\,e_3^2\,e_4^2\,u^2\,v + s_3\,e_2^2\,e_3^3\,u\,v^2 \\
	+  &  s_5\,e_1^2\,e_2\,e_4^3\,u^2\,w + s_6\,e_1\,e_2\,e_3\,e_4\,u\,v\,w + s_9\,e_1\,v\,w^2 = 0,
\end{split}
\end{align}
whose vanishing locus defines the resolved elliptic fibration $Y_4$ \cite{Klevers:2014bqa}.
The coefficients $s_i$ remain unchanged compared to \eqref{eq:coefficients_F11_classes} as functions on the base $B_3$.

On the resolved fibration $\pi: Y_4 \rightarrow B_3$, we can determine the most generic vertical $G_4$-flux \textit{base-independently} \cite{Cvetic:2015txa,Lin:2015qsa,Lin:2016vus,Cvetic:2013uta,MayorgaPena:2017eda}.
It is parametrized by $a \in \mathbb{Q}$ and $\omega \in H^{1,1}(B_3)$ and can be written as
\begin{align}\label{eq:generic_flux}
	G_4 (a , \, \omega) = a\,G_4^a + \pi^*\omega \wedge \sigma \, ,
\end{align}
where $\sigma$ is the $(1,1)$-form Poincar\'{e}-dual to the Shioda-divisor associated with the $U(1)$ (see \cite{Klevers:2014bqa} for its explicit expression).
The $(2,2)$-form $G_4^a$ is 
\begin{align}
	\begin{split}
		G_4^a = \, & [e_4] \wedge ([e_4] + \pi^*[s_6]) \\
		+ \, & \frac{[e_1] \wedge \pi^*[s_9]}{2} + \frac{\pi^*[s_3] \wedge ([e_2] + 2[u])}{3} \, ,
	\end{split}
\end{align}
where $[x]$ denotes the $(1,1)$-form Poincar\'{e}-dual to the divisor $\{x=0\} \subset Y_4$.

In the following, we show how for the restriction ${\cal S}_{7,9} = \KB$, the flux satisfies the necessary consistency conditions.
This is based on topological computation methods developed in \cite{Lin:2015qsa,Lin:2016vus,Cvetic:2013uta,MayorgaPena:2017eda}.
These allow to reduce the relevant integral expressions in the fibration $Y_4$ to intersection numbers on the base $B_3$.

First, we reduce the masslessness condition \eqref{eq:U1-D-term} for $U(1)_Y$ with these methods, which becomes
\begin{align}
	\forall \eta \in H^{1,1}(B_3): \quad \int_{B_3} \KB \wedge \eta \wedge (5 \, \omega + a \, \KB) \stackrel{!}{=} 0 \, .
\end{align}
This can always be satisfied if $\omega = - 5\KB / a$, regardless of the base.
Inserting this into the condition \eqref{eq:chirality_formula} for three chiral families, we then find
\begin{align}
	\chi({\bf R}) = -\frac{a}{5}\int_{B_3} \KB \wedge \KB \wedge \KB =: -\frac{a}{5}\,\KB^3 \, .
\end{align}
For exactly three families, we thus need $a = -15/\KB^3$.
This fixes the flux parameters $(a,\omega)$ completely.

A necessary condition for this flux to satisfy the quantization condition \eqref{eq:quantization_condition} is for the following integrals to be integers:
\begin{align}
\begin{split}
	& \int_{Y_4} \left( G_4  + \frac12 c_2(Y_4) \right) \wedge \text{PD}(\gamma_{\bf R}) \, , \\
	& \int_{Y_4} \left( G_4  + \frac12 c_2(Y_4) \right) \wedge \text{PD}(D_1 \cdot D_2) \, ,
\end{split}
\end{align}
where the integrands contain the Poincar\'{e}-duals (PD) of all matter surfaces $\gamma_{\bf R}$ as well as intersection products $D_1 \cdot D_2$ of all integral divisors $D_i$.
For the smooth elliptic fibrations, $D_i$ are known to be independent fibral divisors $\{u=0\}, \{v=0\}, \{w=0\}$, $\{e_i =0\}$, and $\pi^{-1}(D_B)$ for $D_B$ any integral divisor of $B_3$.
The data for $\gamma_{\bf R}$ and $c_2(Y_4)$ can be found in \cite{Klevers:2014bqa,Cvetic:2015txa}.
After reducing these integrals to intersection numbers on $B_3$, the only non-manifestly integer quantities are
\begin{align}
	& \int_{B_3} \frac{c_2(B)\wedge \KB}{2} \, , \quad \frac{\KB^3}{2} \, , \\
	\text{and } \,\, & \int_{B_3} \frac{\alpha \wedge (c_2(B_3) + \KB^2)}{2} \, \text{ with } \, \alpha \in H^{1,1}(B_3,\bbZ)\, , \notag
\end{align}
where $c_2(B_3)$ is the second Chern class of $B_3$.
For smooth threefolds $B_3$ that appear as a base of a smooth elliptic Calabi--Yau fourfold, it is known \cite{Collinucci:2010gz} that $\int_{B_3} c_2(B_3) \wedge \KB = 24$ and that $c_2(B_3) + \KB^2$ is an \textit{even} class.

The remaining condition of $\KB^3$ being even is covered by the D3-tadpole \eqref{eq:D3-tadpole_general}, which with the above simplifications becomes
\begin{align}
	n_\text{D3} = 12 + \frac{5}{8}\,\KB^3 - \frac{45}{2\,\KB^3} .
\end{align}
Since this is integer if and only if $\KB^3$ is even, all necessary consistency conditions on $G_4$ are satisfied if $n_\text{D3} \in \bbZ_{\geq 0}$.

\newpage

\bibliography{references}

\begin{thebibliography}{57}%
\makeatletter
\providecommand \@ifxundefined [1]{%
 \@ifx{#1\undefined}
}%
\providecommand \@ifnum [1]{%
 \ifnum #1\expandafter \@firstoftwo
 \else \expandafter \@secondoftwo
 \fi
}%
\providecommand \@ifx [1]{%
 \ifx #1\expandafter \@firstoftwo
 \else \expandafter \@secondoftwo
 \fi
}%
\providecommand \natexlab [1]{#1}%
\providecommand \enquote  [1]{``#1''}%
\providecommand \bibnamefont  [1]{#1}%
\providecommand \bibfnamefont [1]{#1}%
\providecommand \citenamefont [1]{#1}%
\providecommand \href@noop [0]{\@secondoftwo}%
\providecommand \href [0]{\begingroup \@sanitize@url \@href}%
\providecommand \@href[1]{\@@startlink{#1}\@@href}%
\providecommand \@@href[1]{\endgroup#1\@@endlink}%
\providecommand \@sanitize@url [0]{\catcode `\\12\catcode `\$12\catcode
  `\&12\catcode `\#12\catcode `\^12\catcode `\_12\catcode `\%12\relax}%
\providecommand \@@startlink[1]{}%
\providecommand \@@endlink[0]{}%
\providecommand \url  [0]{\begingroup\@sanitize@url \@url }%
\providecommand \@url [1]{\endgroup\@href {#1}{\urlprefix }}%
\providecommand \urlprefix  [0]{URL }%
\providecommand \Eprint [0]{\href }%
\providecommand \doibase [0]{http://dx.doi.org/}%
\providecommand \selectlanguage [0]{\@gobble}%
\providecommand \bibinfo  [0]{\@secondoftwo}%
\providecommand \bibfield  [0]{\@secondoftwo}%
\providecommand \translation [1]{[#1]}%
\providecommand \BibitemOpen [0]{}%
\providecommand \bibitemStop [0]{}%
\providecommand \bibitemNoStop [0]{.\EOS\space}%
\providecommand \EOS [0]{\spacefactor3000\relax}%
\providecommand \BibitemShut  [1]{\csname bibitem#1\endcsname}%
\let\auto@bib@innerbib\@empty
\bibitem [{\citenamefont {Vafa}(1996)}]{Vafa:1996xn}%
  \BibitemOpen
  \bibfield  {author} {\bibinfo {author} {\bibfnamefont {C.}~\bibnamefont
  {Vafa}},\ }\href {\doibase 10.1016/0550-3213(96)00172-1} {\bibfield
  {journal} {\bibinfo  {journal} {Nucl. Phys.}\ }\textbf {\bibinfo {volume}
  {B469}},\ \bibinfo {pages} {403} (\bibinfo {year} {1996})},\ \Eprint
  {http://arxiv.org/abs/hep-th/9602022} {arXiv:hep-th/9602022 [hep-th]}
  \BibitemShut {NoStop}%
\bibitem [{\citenamefont {Weigand}(2018)}]{Weigand:2018rez}%
  \BibitemOpen
  \bibfield  {author} {\bibinfo {author} {\bibfnamefont {T.}~\bibnamefont
  {Weigand}},\ }\href@noop {} {\  (\bibinfo {year} {2018})},\ \Eprint
  {http://arxiv.org/abs/1806.01854} {arXiv:1806.01854 [hep-th]} \BibitemShut
  {NoStop}%
\bibitem [{\citenamefont {Cveti{\v c}}\ and\ \citenamefont
  {Lin}(2018{\natexlab{a}})}]{Cvetic:2018bni}%
  \BibitemOpen
  \bibfield  {author} {\bibinfo {author} {\bibfnamefont {M.}~\bibnamefont
  {Cveti{\v c}}}\ and\ \bibinfo {author} {\bibfnamefont {L.}~\bibnamefont
  {Lin}},\ }\href {\doibase 10.22323/1.305.0020} {\bibfield  {journal}
  {\bibinfo  {journal} {PoS}\ }\textbf {\bibinfo {volume} {TASI2017}},\
  \bibinfo {pages} {020} (\bibinfo {year} {2018}{\natexlab{a}})},\ \Eprint
  {http://arxiv.org/abs/1809.00012} {arXiv:1809.00012 [hep-th]} \BibitemShut
  {NoStop}%
\bibitem [{\citenamefont {Ashok}\ and\ \citenamefont
  {Douglas}(2004)}]{Ashok:2003gk}%
  \BibitemOpen
  \bibfield  {author} {\bibinfo {author} {\bibfnamefont {S.}~\bibnamefont
  {Ashok}}\ and\ \bibinfo {author} {\bibfnamefont {M.~R.}\ \bibnamefont
  {Douglas}},\ }\href {\doibase 10.1088/1126-6708/2004/01/060} {\bibfield
  {journal} {\bibinfo  {journal} {JHEP}\ }\textbf {\bibinfo {volume} {01}},\
  \bibinfo {pages} {060} (\bibinfo {year} {2004})},\ \Eprint
  {http://arxiv.org/abs/hep-th/0307049} {arXiv:hep-th/0307049 [hep-th]}
  \BibitemShut {NoStop}%
\bibitem [{\citenamefont {Taylor}\ and\ \citenamefont
  {Wang}(2015)}]{Taylor:2015xtz}%
  \BibitemOpen
  \bibfield  {author} {\bibinfo {author} {\bibfnamefont {W.}~\bibnamefont
  {Taylor}}\ and\ \bibinfo {author} {\bibfnamefont {Y.-N.}\ \bibnamefont
  {Wang}},\ }\href {\doibase 10.1007/JHEP12(2015)164} {\bibfield  {journal}
  {\bibinfo  {journal} {JHEP}\ }\textbf {\bibinfo {volume} {12}},\ \bibinfo
  {pages} {164} (\bibinfo {year} {2015})},\ \Eprint
  {http://arxiv.org/abs/1511.03209} {arXiv:1511.03209 [hep-th]} \BibitemShut
  {NoStop}%
\bibitem [{\citenamefont {Halverson}\ \emph {et~al.}(2017)\citenamefont
  {Halverson}, \citenamefont {Long},\ and\ \citenamefont
  {Sung}}]{Halverson:2017ffz}%
  \BibitemOpen
  \bibfield  {author} {\bibinfo {author} {\bibfnamefont {J.}~\bibnamefont
  {Halverson}}, \bibinfo {author} {\bibfnamefont {C.}~\bibnamefont {Long}}, \
  and\ \bibinfo {author} {\bibfnamefont {B.}~\bibnamefont {Sung}},\ }\href
  {\doibase 10.1103/PhysRevD.96.126006} {\bibfield  {journal} {\bibinfo
  {journal} {Phys. Rev.}\ }\textbf {\bibinfo {volume} {D96}},\ \bibinfo {pages}
  {126006} (\bibinfo {year} {2017})},\ \Eprint
  {http://arxiv.org/abs/1706.02299} {arXiv:1706.02299 [hep-th]} \BibitemShut
  {NoStop}%
\bibitem [{\citenamefont {Blumenhagen}\ \emph
  {et~al.}(2005{\natexlab{a}})\citenamefont {Blumenhagen}, \citenamefont
  {Gmeiner}, \citenamefont {Honecker}, \citenamefont {L{\"u}st},\ and\
  \citenamefont {Weigand}}]{Blumenhagen:2004xx}%
  \BibitemOpen
  \bibfield  {author} {\bibinfo {author} {\bibfnamefont {R.}~\bibnamefont
  {Blumenhagen}}, \bibinfo {author} {\bibfnamefont {F.}~\bibnamefont
  {Gmeiner}}, \bibinfo {author} {\bibfnamefont {G.}~\bibnamefont {Honecker}},
  \bibinfo {author} {\bibfnamefont {D.}~\bibnamefont {L{\"u}st}}, \ and\
  \bibinfo {author} {\bibfnamefont {T.}~\bibnamefont {Weigand}},\ }\href
  {\doibase 10.1016/j.nuclphysb.2005.02.005} {\bibfield  {journal} {\bibinfo
  {journal} {Nucl. Phys.}\ }\textbf {\bibinfo {volume} {B713}},\ \bibinfo
  {pages} {83} (\bibinfo {year} {2005}{\natexlab{a}})},\ \Eprint
  {http://arxiv.org/abs/hep-th/0411173} {arXiv:hep-th/0411173 [hep-th]}
  \BibitemShut {NoStop}%
\bibitem [{\citenamefont {Gmeiner}\ \emph {et~al.}(2006)\citenamefont
  {Gmeiner}, \citenamefont {Blumenhagen}, \citenamefont {Honecker},
  \citenamefont {L{\"u}st},\ and\ \citenamefont {Weigand}}]{Gmeiner:2005vz}%
  \BibitemOpen
  \bibfield  {author} {\bibinfo {author} {\bibfnamefont {F.}~\bibnamefont
  {Gmeiner}}, \bibinfo {author} {\bibfnamefont {R.}~\bibnamefont
  {Blumenhagen}}, \bibinfo {author} {\bibfnamefont {G.}~\bibnamefont
  {Honecker}}, \bibinfo {author} {\bibfnamefont {D.}~\bibnamefont {L{\"u}st}},
  \ and\ \bibinfo {author} {\bibfnamefont {T.}~\bibnamefont {Weigand}},\ }\href
  {\doibase 10.1088/1126-6708/2006/01/004} {\bibfield  {journal} {\bibinfo
  {journal} {JHEP}\ }\textbf {\bibinfo {volume} {01}},\ \bibinfo {pages} {004}
  (\bibinfo {year} {2006})},\ \Eprint {http://arxiv.org/abs/hep-th/0510170}
  {arXiv:hep-th/0510170 [hep-th]} \BibitemShut {NoStop}%
\bibitem [{\citenamefont {Anderson}\ \emph {et~al.}(2012)\citenamefont
  {Anderson}, \citenamefont {Gray}, \citenamefont {Lukas},\ and\ \citenamefont
  {Palti}}]{Anderson:2012yf}%
  \BibitemOpen
  \bibfield  {author} {\bibinfo {author} {\bibfnamefont {L.~B.}\ \bibnamefont
  {Anderson}}, \bibinfo {author} {\bibfnamefont {J.}~\bibnamefont {Gray}},
  \bibinfo {author} {\bibfnamefont {A.}~\bibnamefont {Lukas}}, \ and\ \bibinfo
  {author} {\bibfnamefont {E.}~\bibnamefont {Palti}},\ }\href {\doibase
  10.1007/JHEP06(2012)113} {\bibfield  {journal} {\bibinfo  {journal} {JHEP}\
  }\textbf {\bibinfo {volume} {06}},\ \bibinfo {pages} {113} (\bibinfo {year}
  {2012})},\ \Eprint {http://arxiv.org/abs/1202.1757} {arXiv:1202.1757
  [hep-th]} \BibitemShut {NoStop}%
\bibitem [{\citenamefont {Anderson}\ \emph {et~al.}(2014)\citenamefont
  {Anderson}, \citenamefont {Constantin}, \citenamefont {Gray}, \citenamefont
  {Lukas},\ and\ \citenamefont {Palti}}]{Anderson:2013xka}%
  \BibitemOpen
  \bibfield  {author} {\bibinfo {author} {\bibfnamefont {L.~B.}\ \bibnamefont
  {Anderson}}, \bibinfo {author} {\bibfnamefont {A.}~\bibnamefont
  {Constantin}}, \bibinfo {author} {\bibfnamefont {J.}~\bibnamefont {Gray}},
  \bibinfo {author} {\bibfnamefont {A.}~\bibnamefont {Lukas}}, \ and\ \bibinfo
  {author} {\bibfnamefont {E.}~\bibnamefont {Palti}},\ }\href {\doibase
  10.1007/JHEP01(2014)047} {\bibfield  {journal} {\bibinfo  {journal} {JHEP}\
  }\textbf {\bibinfo {volume} {01}},\ \bibinfo {pages} {047} (\bibinfo {year}
  {2014})},\ \Eprint {http://arxiv.org/abs/1307.4787} {arXiv:1307.4787
  [hep-th]} \BibitemShut {NoStop}%
\bibitem [{\citenamefont {Constantin}\ \emph {et~al.}(2019)\citenamefont
  {Constantin}, \citenamefont {He},\ and\ \citenamefont
  {Lukas}}]{Constantin:2018xkj}%
  \BibitemOpen
  \bibfield  {author} {\bibinfo {author} {\bibfnamefont {A.}~\bibnamefont
  {Constantin}}, \bibinfo {author} {\bibfnamefont {Y.-H.}\ \bibnamefont {He}},
  \ and\ \bibinfo {author} {\bibfnamefont {A.}~\bibnamefont {Lukas}},\ }\href
  {\doibase 10.1016/j.physletb.2019.03.048} {\bibfield  {journal} {\bibinfo
  {journal} {Phys. Lett.}\ }\textbf {\bibinfo {volume} {B792}},\ \bibinfo
  {pages} {258} (\bibinfo {year} {2019})},\ \Eprint
  {http://arxiv.org/abs/1810.00444} {arXiv:1810.00444 [hep-th]} \BibitemShut
  {NoStop}%
\bibitem [{\citenamefont {Cveti{\v c}}\ \emph
  {et~al.}(2001{\natexlab{a}})\citenamefont {Cveti{\v c}}, \citenamefont
  {Shiu},\ and\ \citenamefont {Uranga}}]{Cvetic:2001nr}%
  \BibitemOpen
  \bibfield  {author} {\bibinfo {author} {\bibfnamefont {M.}~\bibnamefont
  {Cveti{\v c}}}, \bibinfo {author} {\bibfnamefont {G.}~\bibnamefont {Shiu}}, \
  and\ \bibinfo {author} {\bibfnamefont {A.~M.}\ \bibnamefont {Uranga}},\
  }\href {\doibase 10.1016/S0550-3213(01)00427-8} {\bibfield  {journal}
  {\bibinfo  {journal} {Nucl. Phys.}\ }\textbf {\bibinfo {volume} {B615}},\
  \bibinfo {pages} {3} (\bibinfo {year} {2001}{\natexlab{a}})},\ \Eprint
  {http://arxiv.org/abs/hep-th/0107166} {arXiv:hep-th/0107166 [hep-th]}
  \BibitemShut {NoStop}%
\bibitem [{\citenamefont {Cveti{\v c}}\ \emph
  {et~al.}(2001{\natexlab{b}})\citenamefont {Cveti{\v c}}, \citenamefont
  {Shiu},\ and\ \citenamefont {Uranga}}]{Cvetic:2001tj}%
  \BibitemOpen
  \bibfield  {author} {\bibinfo {author} {\bibfnamefont {M.}~\bibnamefont
  {Cveti{\v c}}}, \bibinfo {author} {\bibfnamefont {G.}~\bibnamefont {Shiu}}, \
  and\ \bibinfo {author} {\bibfnamefont {A.~M.}\ \bibnamefont {Uranga}},\
  }\href {\doibase 10.1103/PhysRevLett.87.201801} {\bibfield  {journal}
  {\bibinfo  {journal} {Phys. Rev. Lett.}\ }\textbf {\bibinfo {volume} {87}},\
  \bibinfo {pages} {201801} (\bibinfo {year} {2001}{\natexlab{b}})},\ \Eprint
  {http://arxiv.org/abs/hep-th/0107143} {arXiv:hep-th/0107143 [hep-th]}
  \BibitemShut {NoStop}%
\bibitem [{\citenamefont {Blumenhagen}\ \emph
  {et~al.}(2005{\natexlab{b}})\citenamefont {Blumenhagen}, \citenamefont
  {Cveti{\v c}}, \citenamefont {Langacker},\ and\ \citenamefont
  {Shiu}}]{Blumenhagen:2005mu}%
  \BibitemOpen
  \bibfield  {author} {\bibinfo {author} {\bibfnamefont {R.}~\bibnamefont
  {Blumenhagen}}, \bibinfo {author} {\bibfnamefont {M.}~\bibnamefont {Cveti{\v
  c}}}, \bibinfo {author} {\bibfnamefont {P.}~\bibnamefont {Langacker}}, \ and\
  \bibinfo {author} {\bibfnamefont {G.}~\bibnamefont {Shiu}},\ }\href {\doibase
  10.1146/annurev.nucl.55.090704.151541} {\bibfield  {journal} {\bibinfo
  {journal} {Ann. Rev. Nucl. Part. Sci.}\ }\textbf {\bibinfo {volume} {55}},\
  \bibinfo {pages} {71} (\bibinfo {year} {2005}{\natexlab{b}})},\ \Eprint
  {http://arxiv.org/abs/hep-th/0502005} {arXiv:hep-th/0502005 [hep-th]}
  \BibitemShut {NoStop}%
\bibitem [{\citenamefont {Braun}\ \emph {et~al.}(2005)\citenamefont {Braun},
  \citenamefont {He}, \citenamefont {Ovrut},\ and\ \citenamefont
  {Pantev}}]{Braun:2005ux}%
  \BibitemOpen
  \bibfield  {author} {\bibinfo {author} {\bibfnamefont {V.}~\bibnamefont
  {Braun}}, \bibinfo {author} {\bibfnamefont {Y.-H.}\ \bibnamefont {He}},
  \bibinfo {author} {\bibfnamefont {B.~A.}\ \bibnamefont {Ovrut}}, \ and\
  \bibinfo {author} {\bibfnamefont {T.}~\bibnamefont {Pantev}},\ }\href
  {\doibase 10.1016/j.physletb.2005.05.007} {\bibfield  {journal} {\bibinfo
  {journal} {Phys. Lett.}\ }\textbf {\bibinfo {volume} {B618}},\ \bibinfo
  {pages} {252} (\bibinfo {year} {2005})},\ \Eprint
  {http://arxiv.org/abs/hep-th/0501070} {arXiv:hep-th/0501070 [hep-th]}
  \BibitemShut {NoStop}%
\bibitem [{\citenamefont {Braun}\ \emph {et~al.}(2006)\citenamefont {Braun},
  \citenamefont {He}, \citenamefont {Ovrut},\ and\ \citenamefont
  {Pantev}}]{Braun:2005nv}%
  \BibitemOpen
  \bibfield  {author} {\bibinfo {author} {\bibfnamefont {V.}~\bibnamefont
  {Braun}}, \bibinfo {author} {\bibfnamefont {Y.-H.}\ \bibnamefont {He}},
  \bibinfo {author} {\bibfnamefont {B.~A.}\ \bibnamefont {Ovrut}}, \ and\
  \bibinfo {author} {\bibfnamefont {T.}~\bibnamefont {Pantev}},\ }\href
  {\doibase 10.1088/1126-6708/2006/05/043} {\bibfield  {journal} {\bibinfo
  {journal} {JHEP}\ }\textbf {\bibinfo {volume} {05}},\ \bibinfo {pages} {043}
  (\bibinfo {year} {2006})},\ \Eprint {http://arxiv.org/abs/hep-th/0512177}
  {arXiv:hep-th/0512177 [hep-th]} \BibitemShut {NoStop}%
\bibitem [{\citenamefont {Bouchard}\ and\ \citenamefont
  {Donagi}(2006)}]{Bouchard:2005ag}%
  \BibitemOpen
  \bibfield  {author} {\bibinfo {author} {\bibfnamefont {V.}~\bibnamefont
  {Bouchard}}\ and\ \bibinfo {author} {\bibfnamefont {R.}~\bibnamefont
  {Donagi}},\ }\href {\doibase 10.1016/j.physletb.2005.12.042} {\bibfield
  {journal} {\bibinfo  {journal} {Phys. Lett.}\ }\textbf {\bibinfo {volume}
  {B633}},\ \bibinfo {pages} {783} (\bibinfo {year} {2006})},\ \Eprint
  {http://arxiv.org/abs/hep-th/0512149} {arXiv:hep-th/0512149 [hep-th]}
  \BibitemShut {NoStop}%
\bibitem [{\citenamefont {Bouchard}\ \emph {et~al.}(2006)\citenamefont
  {Bouchard}, \citenamefont {Cveti{\v c}},\ and\ \citenamefont
  {Donagi}}]{Bouchard:2006dn}%
  \BibitemOpen
  \bibfield  {author} {\bibinfo {author} {\bibfnamefont {V.}~\bibnamefont
  {Bouchard}}, \bibinfo {author} {\bibfnamefont {M.}~\bibnamefont {Cveti{\v
  c}}}, \ and\ \bibinfo {author} {\bibfnamefont {R.}~\bibnamefont {Donagi}},\
  }\href {\doibase 10.1016/j.nuclphysb.2006.03.032} {\bibfield  {journal}
  {\bibinfo  {journal} {Nucl. Phys.}\ }\textbf {\bibinfo {volume} {B745}},\
  \bibinfo {pages} {62} (\bibinfo {year} {2006})},\ \Eprint
  {http://arxiv.org/abs/hep-th/0602096} {arXiv:hep-th/0602096 [hep-th]}
  \BibitemShut {NoStop}%
\bibitem [{\citenamefont {Anderson}\ \emph {et~al.}(2007)\citenamefont
  {Anderson}, \citenamefont {He},\ and\ \citenamefont
  {Lukas}}]{Anderson:2007nc}%
  \BibitemOpen
  \bibfield  {author} {\bibinfo {author} {\bibfnamefont {L.~B.}\ \bibnamefont
  {Anderson}}, \bibinfo {author} {\bibfnamefont {Y.-H.}\ \bibnamefont {He}}, \
  and\ \bibinfo {author} {\bibfnamefont {A.}~\bibnamefont {Lukas}},\ }\href
  {\doibase 10.1088/1126-6708/2007/07/049} {\bibfield  {journal} {\bibinfo
  {journal} {JHEP}\ }\textbf {\bibinfo {volume} {07}},\ \bibinfo {pages} {049}
  (\bibinfo {year} {2007})},\ \Eprint {http://arxiv.org/abs/hep-th/0702210}
  {arXiv:hep-th/0702210 [HEP-TH]} \BibitemShut {NoStop}%
\bibitem [{\citenamefont {Anderson}\ \emph {et~al.}(2010)\citenamefont
  {Anderson}, \citenamefont {Gray}, \citenamefont {He},\ and\ \citenamefont
  {Lukas}}]{Anderson:2009mh}%
  \BibitemOpen
  \bibfield  {author} {\bibinfo {author} {\bibfnamefont {L.~B.}\ \bibnamefont
  {Anderson}}, \bibinfo {author} {\bibfnamefont {J.}~\bibnamefont {Gray}},
  \bibinfo {author} {\bibfnamefont {Y.-H.}\ \bibnamefont {He}}, \ and\ \bibinfo
  {author} {\bibfnamefont {A.}~\bibnamefont {Lukas}},\ }\href {\doibase
  10.1007/JHEP02(2010)054} {\bibfield  {journal} {\bibinfo  {journal} {JHEP}\
  }\textbf {\bibinfo {volume} {02}},\ \bibinfo {pages} {054} (\bibinfo {year}
  {2010})},\ \Eprint {http://arxiv.org/abs/0911.1569} {arXiv:0911.1569
  [hep-th]} \BibitemShut {NoStop}%
\bibitem [{\citenamefont {Cveti{\v c}}\ \emph {et~al.}(2015)\citenamefont
  {Cveti{\v c}}, \citenamefont {Klevers}, \citenamefont {Pe{\~ n}a},
  \citenamefont {Oehlmann},\ and\ \citenamefont {Reuter}}]{Cvetic:2015txa}%
  \BibitemOpen
  \bibfield  {author} {\bibinfo {author} {\bibfnamefont {M.}~\bibnamefont
  {Cveti{\v c}}}, \bibinfo {author} {\bibfnamefont {D.}~\bibnamefont
  {Klevers}}, \bibinfo {author} {\bibfnamefont {D.~K.~M.}\ \bibnamefont {Pe{\~
  n}a}}, \bibinfo {author} {\bibfnamefont {P.-K.}\ \bibnamefont {Oehlmann}}, \
  and\ \bibinfo {author} {\bibfnamefont {J.}~\bibnamefont {Reuter}},\ }\href
  {\doibase 10.1007/JHEP08(2015)087} {\bibfield  {journal} {\bibinfo  {journal}
  {JHEP}\ }\textbf {\bibinfo {volume} {08}},\ \bibinfo {pages} {087} (\bibinfo
  {year} {2015})},\ \Eprint {http://arxiv.org/abs/1503.02068} {arXiv:1503.02068
  [hep-th]} \BibitemShut {NoStop}%
\bibitem [{\citenamefont {Cveti{\v c}}\ \emph {et~al.}(2018)\citenamefont
  {Cveti{\v c}}, \citenamefont {Lin}, \citenamefont {Liu},\ and\ \citenamefont
  {Oehlmann}}]{Cvetic:2018ryq}%
  \BibitemOpen
  \bibfield  {author} {\bibinfo {author} {\bibfnamefont {M.}~\bibnamefont
  {Cveti{\v c}}}, \bibinfo {author} {\bibfnamefont {L.}~\bibnamefont {Lin}},
  \bibinfo {author} {\bibfnamefont {M.}~\bibnamefont {Liu}}, \ and\ \bibinfo
  {author} {\bibfnamefont {P.-K.}\ \bibnamefont {Oehlmann}},\ }\href {\doibase
  10.1007/JHEP09(2018)089} {\bibfield  {journal} {\bibinfo  {journal} {JHEP}\
  }\textbf {\bibinfo {volume} {09}},\ \bibinfo {pages} {089} (\bibinfo {year}
  {2018})},\ \Eprint {http://arxiv.org/abs/1807.01320} {arXiv:1807.01320
  [hep-th]} \BibitemShut {NoStop}%
\bibitem [{\citenamefont {Halverson}\ and\ \citenamefont
  {Tian}(2017)}]{Halverson:2016tve}%
  \BibitemOpen
  \bibfield  {author} {\bibinfo {author} {\bibfnamefont {J.}~\bibnamefont
  {Halverson}}\ and\ \bibinfo {author} {\bibfnamefont {J.}~\bibnamefont
  {Tian}},\ }\href {\doibase 10.1103/PhysRevD.95.026005} {\bibfield  {journal}
  {\bibinfo  {journal} {Phys. Rev.}\ }\textbf {\bibinfo {volume} {D95}},\
  \bibinfo {pages} {026005} (\bibinfo {year} {2017})},\ \Eprint
  {http://arxiv.org/abs/1610.08864} {arXiv:1610.08864 [hep-th]} \BibitemShut
  {NoStop}%
\bibitem [{\citenamefont {Carifio}\ \emph {et~al.}(2017)\citenamefont
  {Carifio}, \citenamefont {Halverson}, \citenamefont {Krioukov},\ and\
  \citenamefont {Nelson}}]{Carifio:2017bov}%
  \BibitemOpen
  \bibfield  {author} {\bibinfo {author} {\bibfnamefont {J.}~\bibnamefont
  {Carifio}}, \bibinfo {author} {\bibfnamefont {J.}~\bibnamefont {Halverson}},
  \bibinfo {author} {\bibfnamefont {D.}~\bibnamefont {Krioukov}}, \ and\
  \bibinfo {author} {\bibfnamefont {B.~D.}\ \bibnamefont {Nelson}},\ }\href
  {\doibase 10.1007/JHEP09(2017)157} {\bibfield  {journal} {\bibinfo  {journal}
  {JHEP}\ }\textbf {\bibinfo {volume} {09}},\ \bibinfo {pages} {157} (\bibinfo
  {year} {2017})},\ \Eprint {http://arxiv.org/abs/1707.00655} {arXiv:1707.00655
  [hep-th]} \BibitemShut {NoStop}%
\bibitem [{\citenamefont {Altman}\ \emph {et~al.}(2019)\citenamefont {Altman},
  \citenamefont {Carifio}, \citenamefont {Halverson},\ and\ \citenamefont
  {Nelson}}]{Altman:2018zlc}%
  \BibitemOpen
  \bibfield  {author} {\bibinfo {author} {\bibfnamefont {R.}~\bibnamefont
  {Altman}}, \bibinfo {author} {\bibfnamefont {J.}~\bibnamefont {Carifio}},
  \bibinfo {author} {\bibfnamefont {J.}~\bibnamefont {Halverson}}, \ and\
  \bibinfo {author} {\bibfnamefont {B.~D.}\ \bibnamefont {Nelson}},\ }\href
  {\doibase 10.1007/JHEP03(2019)186} {\bibfield  {journal} {\bibinfo  {journal}
  {JHEP}\ }\textbf {\bibinfo {volume} {03}},\ \bibinfo {pages} {186} (\bibinfo
  {year} {2019})},\ \Eprint {http://arxiv.org/abs/1811.06490} {arXiv:1811.06490
  [hep-th]} \BibitemShut {NoStop}%
\bibitem [{\citenamefont {Klevers}\ \emph {et~al.}(2015)\citenamefont
  {Klevers}, \citenamefont {Mayorga~Pe{\~ n}a}, \citenamefont {Oehlmann},
  \citenamefont {Piragua},\ and\ \citenamefont {Reuter}}]{Klevers:2014bqa}%
  \BibitemOpen
  \bibfield  {author} {\bibinfo {author} {\bibfnamefont {D.}~\bibnamefont
  {Klevers}}, \bibinfo {author} {\bibfnamefont {D.~K.}\ \bibnamefont
  {Mayorga~Pe{\~ n}a}}, \bibinfo {author} {\bibfnamefont {P.-K.}\ \bibnamefont
  {Oehlmann}}, \bibinfo {author} {\bibfnamefont {H.}~\bibnamefont {Piragua}}, \
  and\ \bibinfo {author} {\bibfnamefont {J.}~\bibnamefont {Reuter}},\ }\href
  {\doibase 10.1007/JHEP01(2015)142} {\bibfield  {journal} {\bibinfo  {journal}
  {JHEP}\ }\textbf {\bibinfo {volume} {01}},\ \bibinfo {pages} {142} (\bibinfo
  {year} {2015})},\ \Eprint {http://arxiv.org/abs/1408.4808} {arXiv:1408.4808
  [hep-th]} \BibitemShut {NoStop}%
\bibitem [{\citenamefont {Cveti{\v c}}\ and\ \citenamefont
  {Lin}(2018{\natexlab{b}})}]{Cvetic:2017epq}%
  \BibitemOpen
  \bibfield  {author} {\bibinfo {author} {\bibfnamefont {M.}~\bibnamefont
  {Cveti{\v c}}}\ and\ \bibinfo {author} {\bibfnamefont {L.}~\bibnamefont
  {Lin}},\ }\href {\doibase 10.1007/JHEP01(2018)157} {\bibfield  {journal}
  {\bibinfo  {journal} {JHEP}\ }\textbf {\bibinfo {volume} {01}},\ \bibinfo
  {pages} {157} (\bibinfo {year} {2018}{\natexlab{b}})},\ \Eprint
  {http://arxiv.org/abs/1706.08521} {arXiv:1706.08521 [hep-th]} \BibitemShut
  {NoStop}%
\bibitem [{\citenamefont {Lin}\ \emph {et~al.}(2016)\citenamefont {Lin},
  \citenamefont {Mayrhofer}, \citenamefont {Till},\ and\ \citenamefont
  {Weigand}}]{Lin:2015qsa}%
  \BibitemOpen
  \bibfield  {author} {\bibinfo {author} {\bibfnamefont {L.}~\bibnamefont
  {Lin}}, \bibinfo {author} {\bibfnamefont {C.}~\bibnamefont {Mayrhofer}},
  \bibinfo {author} {\bibfnamefont {O.}~\bibnamefont {Till}}, \ and\ \bibinfo
  {author} {\bibfnamefont {T.}~\bibnamefont {Weigand}},\ }\href {\doibase
  10.1007/JHEP01(2016)098} {\bibfield  {journal} {\bibinfo  {journal} {JHEP}\
  }\textbf {\bibinfo {volume} {01}},\ \bibinfo {pages} {098} (\bibinfo {year}
  {2016})},\ \Eprint {http://arxiv.org/abs/1508.00162} {arXiv:1508.00162
  [hep-th]} \BibitemShut {NoStop}%
\bibitem [{\citenamefont {Lin}\ and\ \citenamefont
  {Weigand}(2016)}]{Lin:2016vus}%
  \BibitemOpen
  \bibfield  {author} {\bibinfo {author} {\bibfnamefont {L.}~\bibnamefont
  {Lin}}\ and\ \bibinfo {author} {\bibfnamefont {T.}~\bibnamefont {Weigand}},\
  }\href {\doibase 10.1016/j.nuclphysb.2016.09.008} {\bibfield  {journal}
  {\bibinfo  {journal} {Nucl. Phys.}\ }\textbf {\bibinfo {volume} {B913}},\
  \bibinfo {pages} {209} (\bibinfo {year} {2016})},\ \Eprint
  {http://arxiv.org/abs/1604.04292} {arXiv:1604.04292 [hep-th]} \BibitemShut
  {NoStop}%
\bibitem [{\citenamefont {Cveti{\v c}}\ \emph {et~al.}(2014)\citenamefont
  {Cveti{\v c}}, \citenamefont {Grassi}, \citenamefont {Klevers},\ and\
  \citenamefont {Piragua}}]{Cvetic:2013uta}%
  \BibitemOpen
  \bibfield  {author} {\bibinfo {author} {\bibfnamefont {M.}~\bibnamefont
  {Cveti{\v c}}}, \bibinfo {author} {\bibfnamefont {A.}~\bibnamefont {Grassi}},
  \bibinfo {author} {\bibfnamefont {D.}~\bibnamefont {Klevers}}, \ and\
  \bibinfo {author} {\bibfnamefont {H.}~\bibnamefont {Piragua}},\ }\href
  {\doibase 10.1007/JHEP04(2014)010} {\bibfield  {journal} {\bibinfo  {journal}
  {JHEP}\ }\textbf {\bibinfo {volume} {04}},\ \bibinfo {pages} {010} (\bibinfo
  {year} {2014})},\ \Eprint {http://arxiv.org/abs/1306.3987} {arXiv:1306.3987
  [hep-th]} \BibitemShut {NoStop}%
\bibitem [{\citenamefont {Mayorga~Pena}\ and\ \citenamefont
  {Valandro}(2018)}]{MayorgaPena:2017eda}%
  \BibitemOpen
  \bibfield  {author} {\bibinfo {author} {\bibfnamefont {D.~K.}\ \bibnamefont
  {Mayorga~Pena}}\ and\ \bibinfo {author} {\bibfnamefont {R.}~\bibnamefont
  {Valandro}},\ }\href {\doibase 10.1007/JHEP03(2018)107} {\bibfield  {journal}
  {\bibinfo  {journal} {JHEP}\ }\textbf {\bibinfo {volume} {03}},\ \bibinfo
  {pages} {107} (\bibinfo {year} {2018})},\ \Eprint
  {http://arxiv.org/abs/1708.09452} {arXiv:1708.09452 [hep-th]} \BibitemShut
  {NoStop}%
\bibitem [{\citenamefont {Dasgupta}\ \emph {et~al.}(1999)\citenamefont
  {Dasgupta}, \citenamefont {Rajesh},\ and\ \citenamefont
  {Sethi}}]{Dasgupta:1999ss}%
  \BibitemOpen
  \bibfield  {author} {\bibinfo {author} {\bibfnamefont {K.}~\bibnamefont
  {Dasgupta}}, \bibinfo {author} {\bibfnamefont {G.}~\bibnamefont {Rajesh}}, \
  and\ \bibinfo {author} {\bibfnamefont {S.}~\bibnamefont {Sethi}},\ }\href
  {\doibase 10.1088/1126-6708/1999/08/023} {\bibfield  {journal} {\bibinfo
  {journal} {JHEP}\ }\textbf {\bibinfo {volume} {08}},\ \bibinfo {pages} {023}
  (\bibinfo {year} {1999})},\ \Eprint {http://arxiv.org/abs/hep-th/9908088}
  {arXiv:hep-th/9908088 [hep-th]} \BibitemShut {NoStop}%
\bibitem [{\citenamefont {Witten}(1997)}]{Witten:1996md}%
  \BibitemOpen
  \bibfield  {author} {\bibinfo {author} {\bibfnamefont {E.}~\bibnamefont
  {Witten}},\ }\href {\doibase 10.1016/S0393-0440(96)00042-3} {\bibfield
  {journal} {\bibinfo  {journal} {J. Geom. Phys.}\ }\textbf {\bibinfo {volume}
  {22}},\ \bibinfo {pages} {1} (\bibinfo {year} {1997})},\ \Eprint
  {http://arxiv.org/abs/hep-th/9609122} {arXiv:hep-th/9609122 [hep-th]}
  \BibitemShut {NoStop}%
\bibitem [{\citenamefont {Collinucci}\ and\ \citenamefont
  {Savelli}(2012)}]{Collinucci:2010gz}%
  \BibitemOpen
  \bibfield  {author} {\bibinfo {author} {\bibfnamefont {A.}~\bibnamefont
  {Collinucci}}\ and\ \bibinfo {author} {\bibfnamefont {R.}~\bibnamefont
  {Savelli}},\ }\href {\doibase 10.1007/JHEP02(2012)015} {\bibfield  {journal}
  {\bibinfo  {journal} {JHEP}\ }\textbf {\bibinfo {volume} {02}},\ \bibinfo
  {pages} {015} (\bibinfo {year} {2012})},\ \Eprint
  {http://arxiv.org/abs/1011.6388} {arXiv:1011.6388 [hep-th]} \BibitemShut
  {NoStop}%
\bibitem [{\citenamefont {Intriligator}\ \emph {et~al.}(2013)\citenamefont
  {Intriligator}, \citenamefont {Jockers}, \citenamefont {Mayr}, \citenamefont
  {Morrison},\ and\ \citenamefont {Plesser}}]{Intriligator:2012ue}%
  \BibitemOpen
  \bibfield  {author} {\bibinfo {author} {\bibfnamefont {K.}~\bibnamefont
  {Intriligator}}, \bibinfo {author} {\bibfnamefont {H.}~\bibnamefont
  {Jockers}}, \bibinfo {author} {\bibfnamefont {P.}~\bibnamefont {Mayr}},
  \bibinfo {author} {\bibfnamefont {D.~R.}\ \bibnamefont {Morrison}}, \ and\
  \bibinfo {author} {\bibfnamefont {M.~R.}\ \bibnamefont {Plesser}},\ }\href
  {\doibase 10.4310/ATMP.2013.v17.n3.a2} {\bibfield  {journal} {\bibinfo
  {journal} {Adv. Theor. Math. Phys.}\ }\textbf {\bibinfo {volume} {17}},\
  \bibinfo {pages} {601} (\bibinfo {year} {2013})},\ \Eprint
  {http://arxiv.org/abs/1203.6662} {arXiv:1203.6662 [hep-th]} \BibitemShut
  {NoStop}%
\bibitem [{\citenamefont {Sethi}\ \emph {et~al.}(1996)\citenamefont {Sethi},
  \citenamefont {Vafa},\ and\ \citenamefont {Witten}}]{Sethi:1996es}%
  \BibitemOpen
  \bibfield  {author} {\bibinfo {author} {\bibfnamefont {S.}~\bibnamefont
  {Sethi}}, \bibinfo {author} {\bibfnamefont {C.}~\bibnamefont {Vafa}}, \ and\
  \bibinfo {author} {\bibfnamefont {E.}~\bibnamefont {Witten}},\ }\href
  {\doibase 10.1016/S0550-3213(96)00483-X} {\bibfield  {journal} {\bibinfo
  {journal} {Nucl. Phys.}\ }\textbf {\bibinfo {volume} {B480}},\ \bibinfo
  {pages} {213} (\bibinfo {year} {1996})},\ \Eprint
  {http://arxiv.org/abs/hep-th/9606122} {arXiv:hep-th/9606122 [hep-th]}
  \BibitemShut {NoStop}%
\bibitem [{\citenamefont {Grimm}(2011)}]{Grimm:2010ks}%
  \BibitemOpen
  \bibfield  {author} {\bibinfo {author} {\bibfnamefont {T.~W.}\ \bibnamefont
  {Grimm}},\ }\href {\doibase 10.1016/j.nuclphysb.2010.11.018} {\bibfield
  {journal} {\bibinfo  {journal} {Nucl. Phys.}\ }\textbf {\bibinfo {volume}
  {B845}},\ \bibinfo {pages} {48} (\bibinfo {year} {2011})},\ \Eprint
  {http://arxiv.org/abs/1008.4133} {arXiv:1008.4133 [hep-th]} \BibitemShut
  {NoStop}%
\bibitem [{\citenamefont {Grimm}\ \emph {et~al.}(2011)\citenamefont {Grimm},
  \citenamefont {Kerstan}, \citenamefont {Palti},\ and\ \citenamefont
  {Weigand}}]{Grimm:2011tb}%
  \BibitemOpen
  \bibfield  {author} {\bibinfo {author} {\bibfnamefont {T.~W.}\ \bibnamefont
  {Grimm}}, \bibinfo {author} {\bibfnamefont {M.}~\bibnamefont {Kerstan}},
  \bibinfo {author} {\bibfnamefont {E.}~\bibnamefont {Palti}}, \ and\ \bibinfo
  {author} {\bibfnamefont {T.}~\bibnamefont {Weigand}},\ }\href {\doibase
  10.1007/JHEP12(2011)004} {\bibfield  {journal} {\bibinfo  {journal} {JHEP}\
  }\textbf {\bibinfo {volume} {12}},\ \bibinfo {pages} {004} (\bibinfo {year}
  {2011})},\ \Eprint {http://arxiv.org/abs/1107.3842} {arXiv:1107.3842
  [hep-th]} \BibitemShut {NoStop}%
\bibitem [{\citenamefont {Park}(2012)}]{Park:2011ji}%
  \BibitemOpen
  \bibfield  {author} {\bibinfo {author} {\bibfnamefont {D.~S.}\ \bibnamefont
  {Park}},\ }\href {\doibase 10.1007/JHEP01(2012)093} {\bibfield  {journal}
  {\bibinfo  {journal} {JHEP}\ }\textbf {\bibinfo {volume} {01}},\ \bibinfo
  {pages} {093} (\bibinfo {year} {2012})},\ \Eprint
  {http://arxiv.org/abs/1111.2351} {arXiv:1111.2351 [hep-th]} \BibitemShut
  {NoStop}%
\bibitem [{\citenamefont {Morrison}\ and\ \citenamefont
  {Park}(2012)}]{Morrison:2012ei}%
  \BibitemOpen
  \bibfield  {author} {\bibinfo {author} {\bibfnamefont {D.~R.}\ \bibnamefont
  {Morrison}}\ and\ \bibinfo {author} {\bibfnamefont {D.~S.}\ \bibnamefont
  {Park}},\ }\href {\doibase 10.1007/JHEP10(2012)128} {\bibfield  {journal}
  {\bibinfo  {journal} {JHEP}\ }\textbf {\bibinfo {volume} {10}},\ \bibinfo
  {pages} {128} (\bibinfo {year} {2012})},\ \Eprint
  {http://arxiv.org/abs/1208.2695} {arXiv:1208.2695 [hep-th]} \BibitemShut
  {NoStop}%
\bibitem [{\citenamefont {Donagi}\ and\ \citenamefont
  {Wijnholt}(2011)}]{Donagi:2008ca}%
  \BibitemOpen
  \bibfield  {author} {\bibinfo {author} {\bibfnamefont {R.}~\bibnamefont
  {Donagi}}\ and\ \bibinfo {author} {\bibfnamefont {M.}~\bibnamefont
  {Wijnholt}},\ }\href {\doibase 10.4310/ATMP.2011.v15.n5.a2} {\bibfield
  {journal} {\bibinfo  {journal} {Adv. Theor. Math. Phys.}\ }\textbf {\bibinfo
  {volume} {15}},\ \bibinfo {pages} {1237} (\bibinfo {year} {2011})},\ \Eprint
  {http://arxiv.org/abs/0802.2969} {arXiv:0802.2969 [hep-th]} \BibitemShut
  {NoStop}%
\bibitem [{\citenamefont {Braun}\ \emph {et~al.}(2012)\citenamefont {Braun},
  \citenamefont {Collinucci},\ and\ \citenamefont {Valandro}}]{Braun:2011zm}%
  \BibitemOpen
  \bibfield  {author} {\bibinfo {author} {\bibfnamefont {A.~P.}\ \bibnamefont
  {Braun}}, \bibinfo {author} {\bibfnamefont {A.}~\bibnamefont {Collinucci}}, \
  and\ \bibinfo {author} {\bibfnamefont {R.}~\bibnamefont {Valandro}},\ }\href
  {\doibase 10.1016/j.nuclphysb.2011.10.034} {\bibfield  {journal} {\bibinfo
  {journal} {Nucl. Phys.}\ }\textbf {\bibinfo {volume} {B856}},\ \bibinfo
  {pages} {129} (\bibinfo {year} {2012})},\ \Eprint
  {http://arxiv.org/abs/1107.5337} {arXiv:1107.5337 [hep-th]} \BibitemShut
  {NoStop}%
\bibitem [{\citenamefont {Marsano}\ and\ \citenamefont {Sch{\"
  a}fer-Nameki}(2011)}]{Marsano:2011hv}%
  \BibitemOpen
  \bibfield  {author} {\bibinfo {author} {\bibfnamefont {J.}~\bibnamefont
  {Marsano}}\ and\ \bibinfo {author} {\bibfnamefont {S.}~\bibnamefont {Sch{\"
  a}fer-Nameki}},\ }\href {\doibase 10.1007/JHEP11(2011)098} {\bibfield
  {journal} {\bibinfo  {journal} {JHEP}\ }\textbf {\bibinfo {volume} {11}},\
  \bibinfo {pages} {098} (\bibinfo {year} {2011})},\ \Eprint
  {http://arxiv.org/abs/1108.1794} {arXiv:1108.1794 [hep-th]} \BibitemShut
  {NoStop}%
\bibitem [{\citenamefont {Krause}\ \emph {et~al.}(2012)\citenamefont {Krause},
  \citenamefont {Mayrhofer},\ and\ \citenamefont {Weigand}}]{Krause:2011xj}%
  \BibitemOpen
  \bibfield  {author} {\bibinfo {author} {\bibfnamefont {S.}~\bibnamefont
  {Krause}}, \bibinfo {author} {\bibfnamefont {C.}~\bibnamefont {Mayrhofer}}, \
  and\ \bibinfo {author} {\bibfnamefont {T.}~\bibnamefont {Weigand}},\ }\href
  {\doibase 10.1016/j.nuclphysb.2011.12.013} {\bibfield  {journal} {\bibinfo
  {journal} {Nucl. Phys.}\ }\textbf {\bibinfo {volume} {B858}},\ \bibinfo
  {pages} {1} (\bibinfo {year} {2012})},\ \Eprint
  {http://arxiv.org/abs/1109.3454} {arXiv:1109.3454 [hep-th]} \BibitemShut
  {NoStop}%
\bibitem [{\citenamefont {Grimm}\ and\ \citenamefont
  {Hayashi}(2012)}]{Grimm:2011fx}%
  \BibitemOpen
  \bibfield  {author} {\bibinfo {author} {\bibfnamefont {T.~W.}\ \bibnamefont
  {Grimm}}\ and\ \bibinfo {author} {\bibfnamefont {H.}~\bibnamefont
  {Hayashi}},\ }\href {\doibase 10.1007/JHEP03(2012)027} {\bibfield  {journal}
  {\bibinfo  {journal} {JHEP}\ }\textbf {\bibinfo {volume} {03}},\ \bibinfo
  {pages} {027} (\bibinfo {year} {2012})},\ \Eprint
  {http://arxiv.org/abs/1111.1232} {arXiv:1111.1232 [hep-th]} \BibitemShut
  {NoStop}%
\bibitem [{\citenamefont {Kreuzer}\ and\ \citenamefont
  {Skarke}(1998)}]{Kreuzer:1998vb}%
  \BibitemOpen
  \bibfield  {author} {\bibinfo {author} {\bibfnamefont {M.}~\bibnamefont
  {Kreuzer}}\ and\ \bibinfo {author} {\bibfnamefont {H.}~\bibnamefont
  {Skarke}},\ }\href {\doibase 10.4310/ATMP.1998.v2.n4.a5} {\bibfield
  {journal} {\bibinfo  {journal} {Adv. Theor. Math. Phys.}\ }\textbf {\bibinfo
  {volume} {2}},\ \bibinfo {pages} {853} (\bibinfo {year} {1998})},\ \Eprint
  {http://arxiv.org/abs/hep-th/9805190} {arXiv:hep-th/9805190 [hep-th]}
  \BibitemShut {NoStop}%
\bibitem [{\citenamefont {Stein}\ \emph {et~al.}(2018)\citenamefont {Stein}
  \emph {et~al.}}]{sage}%
  \BibitemOpen
  \bibfield  {author} {\bibinfo {author} {\bibfnamefont {W.}~\bibnamefont
  {Stein}} \emph {et~al.},\ }\href@noop {} {\emph {\bibinfo {title} {{S}age
  {M}athematics {S}oftware ({V}ersion 8.4)}}},\ \bibinfo {organization} {The
  Sage Development Team} (\bibinfo {year} {2018}),\ \bibinfo {note} {{\tt
  http://www.sagemath.org}}\BibitemShut {NoStop}%
\bibitem [{\citenamefont {Bies}\ \emph {et~al.}(2014)\citenamefont {Bies},
  \citenamefont {Mayrhofer}, \citenamefont {Pehle},\ and\ \citenamefont
  {Weigand}}]{Bies:2014sra}%
  \BibitemOpen
  \bibfield  {author} {\bibinfo {author} {\bibfnamefont {M.}~\bibnamefont
  {Bies}}, \bibinfo {author} {\bibfnamefont {C.}~\bibnamefont {Mayrhofer}},
  \bibinfo {author} {\bibfnamefont {C.}~\bibnamefont {Pehle}}, \ and\ \bibinfo
  {author} {\bibfnamefont {T.}~\bibnamefont {Weigand}},\ }\href@noop {} {\
  (\bibinfo {year} {2014})},\ \Eprint {http://arxiv.org/abs/1402.5144}
  {arXiv:1402.5144 [hep-th]} \BibitemShut {NoStop}%
\bibitem [{\citenamefont {Bies}\ \emph {et~al.}(2017)\citenamefont {Bies},
  \citenamefont {Mayrhofer},\ and\ \citenamefont {Weigand}}]{Bies:2017fam}%
  \BibitemOpen
  \bibfield  {author} {\bibinfo {author} {\bibfnamefont {M.}~\bibnamefont
  {Bies}}, \bibinfo {author} {\bibfnamefont {C.}~\bibnamefont {Mayrhofer}}, \
  and\ \bibinfo {author} {\bibfnamefont {T.}~\bibnamefont {Weigand}},\ }\href
  {\doibase 10.1007/JHEP11(2017)081} {\bibfield  {journal} {\bibinfo  {journal}
  {JHEP}\ }\textbf {\bibinfo {volume} {11}},\ \bibinfo {pages} {081} (\bibinfo
  {year} {2017})},\ \Eprint {http://arxiv.org/abs/1706.04616} {arXiv:1706.04616
  [hep-th]} \BibitemShut {NoStop}%
\bibitem [{\citenamefont {Lin}\ and\ \citenamefont
  {Weigand}(2015)}]{Lin:2014qga}%
  \BibitemOpen
  \bibfield  {author} {\bibinfo {author} {\bibfnamefont {L.}~\bibnamefont
  {Lin}}\ and\ \bibinfo {author} {\bibfnamefont {T.}~\bibnamefont {Weigand}},\
  }\href {\doibase 10.1002/prop.201400072} {\bibfield  {journal} {\bibinfo
  {journal} {Fortsch. Phys.}\ }\textbf {\bibinfo {volume} {63}},\ \bibinfo
  {pages} {55} (\bibinfo {year} {2015})},\ \Eprint
  {http://arxiv.org/abs/1406.6071} {arXiv:1406.6071 [hep-th]} \BibitemShut
  {NoStop}%
\bibitem [{\citenamefont {Bonetti}\ and\ \citenamefont
  {Grimm}(2012)}]{Bonetti:2011mw}%
  \BibitemOpen
  \bibfield  {author} {\bibinfo {author} {\bibfnamefont {F.}~\bibnamefont
  {Bonetti}}\ and\ \bibinfo {author} {\bibfnamefont {T.~W.}\ \bibnamefont
  {Grimm}},\ }\href {\doibase 10.1007/JHEP05(2012)019} {\bibfield  {journal}
  {\bibinfo  {journal} {JHEP}\ }\textbf {\bibinfo {volume} {05}},\ \bibinfo
  {pages} {019} (\bibinfo {year} {2012})},\ \Eprint
  {http://arxiv.org/abs/1112.1082} {arXiv:1112.1082 [hep-th]} \BibitemShut
  {NoStop}%
\bibitem [{\citenamefont {Cveti{\v c}}\ \emph {et~al.}(2013)\citenamefont
  {Cveti{\v c}}, \citenamefont {Grimm},\ and\ \citenamefont
  {Klevers}}]{Cvetic:2012xn}%
  \BibitemOpen
  \bibfield  {author} {\bibinfo {author} {\bibfnamefont {M.}~\bibnamefont
  {Cveti{\v c}}}, \bibinfo {author} {\bibfnamefont {T.~W.}\ \bibnamefont
  {Grimm}}, \ and\ \bibinfo {author} {\bibfnamefont {D.}~\bibnamefont
  {Klevers}},\ }\href {\doibase 10.1007/JHEP02(2013)101} {\bibfield  {journal}
  {\bibinfo  {journal} {JHEP}\ }\textbf {\bibinfo {volume} {02}},\ \bibinfo
  {pages} {101} (\bibinfo {year} {2013})},\ \Eprint
  {http://arxiv.org/abs/1210.6034} {arXiv:1210.6034 [hep-th]} \BibitemShut
  {NoStop}%
\bibitem [{\citenamefont {Kachru}\ \emph {et~al.}(2003)\citenamefont {Kachru},
  \citenamefont {Kallosh}, \citenamefont {Linde},\ and\ \citenamefont
  {Trivedi}}]{Kachru:2003aw}%
  \BibitemOpen
  \bibfield  {author} {\bibinfo {author} {\bibfnamefont {S.}~\bibnamefont
  {Kachru}}, \bibinfo {author} {\bibfnamefont {R.}~\bibnamefont {Kallosh}},
  \bibinfo {author} {\bibfnamefont {A.~D.}\ \bibnamefont {Linde}}, \ and\
  \bibinfo {author} {\bibfnamefont {S.~P.}\ \bibnamefont {Trivedi}},\ }\href
  {\doibase 10.1103/PhysRevD.68.046005} {\bibfield  {journal} {\bibinfo
  {journal} {Phys. Rev.}\ }\textbf {\bibinfo {volume} {D68}},\ \bibinfo {pages}
  {046005} (\bibinfo {year} {2003})},\ \Eprint
  {http://arxiv.org/abs/hep-th/0301240} {arXiv:hep-th/0301240 [hep-th]}
  \BibitemShut {NoStop}%
\bibitem [{\citenamefont {Balasubramanian}\ \emph {et~al.}(2005)\citenamefont
  {Balasubramanian}, \citenamefont {Berglund}, \citenamefont {Conlon},\ and\
  \citenamefont {Quevedo}}]{Balasubramanian:2005zx}%
  \BibitemOpen
  \bibfield  {author} {\bibinfo {author} {\bibfnamefont {V.}~\bibnamefont
  {Balasubramanian}}, \bibinfo {author} {\bibfnamefont {P.}~\bibnamefont
  {Berglund}}, \bibinfo {author} {\bibfnamefont {J.~P.}\ \bibnamefont
  {Conlon}}, \ and\ \bibinfo {author} {\bibfnamefont {F.}~\bibnamefont
  {Quevedo}},\ }\href {\doibase 10.1088/1126-6708/2005/03/007} {\bibfield
  {journal} {\bibinfo  {journal} {JHEP}\ }\textbf {\bibinfo {volume} {03}},\
  \bibinfo {pages} {007} (\bibinfo {year} {2005})},\ \Eprint
  {http://arxiv.org/abs/hep-th/0502058} {arXiv:hep-th/0502058 [hep-th]}
  \BibitemShut {NoStop}%
\bibitem [{\citenamefont {Demirtas}\ \emph {et~al.}(2018)\citenamefont
  {Demirtas}, \citenamefont {Long}, \citenamefont {McAllister},\ and\
  \citenamefont {Stillman}}]{Demirtas:2018akl}%
  \BibitemOpen
  \bibfield  {author} {\bibinfo {author} {\bibfnamefont {M.}~\bibnamefont
  {Demirtas}}, \bibinfo {author} {\bibfnamefont {C.}~\bibnamefont {Long}},
  \bibinfo {author} {\bibfnamefont {L.}~\bibnamefont {McAllister}}, \ and\
  \bibinfo {author} {\bibfnamefont {M.}~\bibnamefont {Stillman}},\ }\href@noop
  {} {\  (\bibinfo {year} {2018})},\ \Eprint {http://arxiv.org/abs/1808.01282}
  {arXiv:1808.01282 [hep-th]} \BibitemShut {NoStop}%
\bibitem [{\citenamefont {Ganor}(1997)}]{Ganor:1996pe}%
  \BibitemOpen
  \bibfield  {author} {\bibinfo {author} {\bibfnamefont {O.~J.}\ \bibnamefont
  {Ganor}},\ }\href {\doibase 10.1016/S0550-3213(97)00311-8} {\bibfield
  {journal} {\bibinfo  {journal} {Nucl. Phys.}\ }\textbf {\bibinfo {volume}
  {B499}},\ \bibinfo {pages} {55} (\bibinfo {year} {1997})},\ \Eprint
  {http://arxiv.org/abs/hep-th/9612077} {arXiv:hep-th/9612077 [hep-th]}
  \BibitemShut {NoStop}%
\bibitem [{\citenamefont {Baumann}\ \emph {et~al.}(2006)\citenamefont
  {Baumann}, \citenamefont {Dymarsky}, \citenamefont {Klebanov}, \citenamefont
  {Maldacena}, \citenamefont {McAllister},\ and\ \citenamefont
  {Murugan}}]{Baumann:2006th}%
  \BibitemOpen
  \bibfield  {author} {\bibinfo {author} {\bibfnamefont {D.}~\bibnamefont
  {Baumann}}, \bibinfo {author} {\bibfnamefont {A.}~\bibnamefont {Dymarsky}},
  \bibinfo {author} {\bibfnamefont {I.~R.}\ \bibnamefont {Klebanov}}, \bibinfo
  {author} {\bibfnamefont {J.~M.}\ \bibnamefont {Maldacena}}, \bibinfo {author}
  {\bibfnamefont {L.~P.}\ \bibnamefont {McAllister}}, \ and\ \bibinfo {author}
  {\bibfnamefont {A.}~\bibnamefont {Murugan}},\ }\href {\doibase
  10.1088/1126-6708/2006/11/031} {\bibfield  {journal} {\bibinfo  {journal}
  {JHEP}\ }\textbf {\bibinfo {volume} {11}},\ \bibinfo {pages} {031} (\bibinfo
  {year} {2006})},\ \Eprint {http://arxiv.org/abs/hep-th/0607050}
  {arXiv:hep-th/0607050 [hep-th]} \BibitemShut {NoStop}%
\end{thebibliography}%

\end{document}